# Functional architecture and global properties of the *Corynebacterium glutamicum* regulatory network: novel insights from a dataset with a high genomic coverage

Julio A. Freyre-González[1,*] and Andreas Tauch[2]

[1]Regulatory Systems Biology Research Group, Evolutionary Genomics Program, Center for Genomics Sciences, Universidad Nacional Autónoma de México. Av. Universidad s/n, Col. Chamilpa, 62210. Cuernavaca, Morelos, México

[2]Centrum für Biotechnologie (CeBiTec), Universität Bielefeld, Universitätsstraße 27, Bielefeld, 33615, Germany.

**\*Corresponding author:** jfreyre@ccg.unam.mx (JAF-G)

**Abstract**

*Corynebacterium glutamicum* is a Gram-positive, anaerobic, rod-shaped soil bacterium able to grow on a diversity of carbon sources like sugars and organic acids. It is a biotechnological relevant organism because of its highly efficient ability to biosynthesize amino acids, such as L-glutamic acid and L-lysine. Here, we reconstructed the most complete *C. glutamicum* regulatory network to date and comprehensively analyzed its global organizational properties, systems-level features and functional architecture. Our analyses show the tremendous power of Abasy Atlas to study the functional organization of regulatory networks. We created two models of the *C. glutamicum* regulatory network: all-evidences (containing both weak and strong supported interactions, genomic coverage = 73%) and strongly-supported (only accounting for strongly supported evidences, genomic coverage = 71%). Using state-of-the-art methodologies, we prove that power-law behaviors truly govern the connectivity and clustering coefficient distributions. We found a non-previously reported circuit motif that we named complex feed-forward motif. We highlighted the importance of feedback loops for the functional architecture, beyond whether they are statistically over-represented or not in the network. We show that the previously reported top-down approach is inadequate to infer the hierarchy governing a regulatory network because feedback bridges different hierarchical layers, and the top-down approach disregards the presence of intermodular genes shaping the integration layer. Our findings all together further support a diamond-shaped, three-layered hierarchy exhibiting some feedback between processing and coordination layers, which is shaped by four classes of systems-level elements: global regulators, locally autonomous modules, basal machinery and intermodular genes.

## 1 Introduction

*Corynebacterium glutamicum* is a Gram-positive, anaerobic, rod-shaped soil bacterium able to grow on a diversity of carbon sources like sugars and organic acids. It is a biotechnological relevant organism because of its highly efficient ability to biosynthesize amino acids, such as L-glutamic acid and L-lysine. In 2003, two research groups, one based in Bielefeld, Germany (Kalinowski et al., 2003) and the other based in Tokyo, Japan (Ikeda and Nakagawa, 2003), reported the genome sequence of the *C. glutamicum* wild-type strain ATCC 13032 independently. Afterwards, a group of researchers at Bielefeld University accepted the challenge of experimentally reconstructing at genome-scale the *C. glutamicum* regulatory network (*C. glutamicum* RN). Three years later, this project produced the database CoryneRegNet (Baumbach et al., 2006), the first electronically-encoded version of the *C. glutamicum* RN. Since then, the regulatory data set has been expanded by numerous new experiments performed by several international research groups and manual literature curation.

Bacterial regulatory networks (bacterial RNs) are responsible for sensing stimuli and environmental cues and responding accordingly. In complex environments, they 'take composite decisions' to prioritize, for example, the transport and catabolism of carbon sources according to the metabolic preferences of each organism. To accomplish this, bacterial RNs, generally composed of thousands of regulatory interactions, must follow well-defined organization principles governing their dynamics. In the last decades of the 20th century, the first levels of gene organization were unveiled as the operon and the regulon. Nevertheless, cumulated evidence has shown that RNs are complex hierarchical-modular structures (Freyre-Gonzalez et al., 2008; Freyre-Gonzalez et al., 2013; Freyre-Gonzalez and Trevino-Quintanilla, 2010; Freyre-Gonzalez et al., 2012; Ibarra-Arellano et al., 2016; Ma et al., 2004a; Resendis-Antonio et al., 2005; Yu and Gerstein, 2006) whose organizational and evolutionary principles, still challenging our understanding, are pivotal for determining the dynamics of the cell.

*C. glutamicum* regulatory data provides an invaluable source of information to improve our understanding of the functional organization of its RN by constructing large-scale models and carrying out analyses based on complex networks theory. Here, we join efforts to reconstruct the most complete *C. glutamicum* RN to date and comprehensively analyze its global organizational properties, systems-level features and functional architecture.

## 2 The natural decomposition approach: A theoretical framework to identify the system-level elements and functional architecture of bacterial regulatory networks

There are different levels of description in models of genetic networks (Bornholdt, 2005). The natural decomposition approach (NDA) is a large-scale modelling approach characterizing the whole wiring of a RN and its global architecture from the intrinsic global properties of a RN. It contrasts with other non-natural methods that use optimization processes and rely on arbitrary parameters that, when modified, can generate a different partition into modules. The NDA defines a mathematical-biological framework providing criteria to identify four classes of systems-level elements shaping RNs and rules to reveal its functional architecture naturally by controlled network decomposition (Figure 1). The four classes of systems-level elements identified by the NDA are global regulators, modular genes shaping functional systems, basal machinery genes, and intermodular genes.

These four classes interrelate in a non-pyramidal, three-layered hierarchy shaping a regulatory functional architecture (Freyre-Gonzalez et al., 2008; Freyre-Gonzalez and Trevino-Quintanilla, 2010; Freyre-Gonzalez et al., 2012) as follows (Figure 2): 1) Global regulators are responsible for coordinating both the 2) basal cell machinery, composed of strictly globally regulated genes and 3) locally autonomous systems (shaped by modular genes), whereas 4) intermodular genes integrate, at promoter level, physiologically disparate module responses eliciting a combinatorial processing of environmental cues.

The NDA is based on two biological premises (Freyre-Gonzalez and Trevino-Quintanilla, 2010; Freyre-Gonzalez et al., 2012): 1) A module is a set of genes cooperating to carry out a particular physiological function, thus conferring different phenotypic traits to the cell. 2) Given the pleiotropic effect of global regulators, they must not belong to modules but rather coordinate them in response to general-interest environmental cues.

Studies have shown that RNs are highly plastic (Price et al., 2007). Despite this plasticity, by applying the NDA and gene orthology analysis it has been found that the organizational principles shaping the functional architecture are maintained in the RNs of phylogenetically distant bacteria whereas genes composing the three layers of the functional architecture (Figure 2) are poorly conserved (Freyre-Gonzalez et al., 2012). Besides, the high predictive power of the NDA has been proven in previous studies by applying it to the phylogenetically distant *Escherichia coli* K-12 MG1655 (Freyre-Gonzalez et al., 2008) and *Bacillus subtilis* 168 (Freyre-Gonzalez et al., 2012) and by comparing it with other methods to identify modules (Freyre-Gonzalez et al., 2013).

Recently, Abasy (**a**cross-**ba**cteria **sy**stems) Atlas (http://abasy.ccg.unam.mx) (Ibarra-Arellano et al., 2016) has been constructed to compute and store the NDA predictions and global properties resulting of a large-scale effort to unravel the functional architectures and systems-level elements across a large range of bacteria. It provides a comprehensive inventory of annotated functional systems (modules), global network properties, and systems-level elements for reconstructed and meta-curated bacterial RNs, including pathogenic and biotechnologically relevant organisms. Besides, Abasy Atlas provides an automated pipeline to compute a large set of global properties characterizing bacterial RNs.

## 3    Construction of the *C. glutamicum* regulatory network models

The *C. glutamicum* RN models here reported largely derived from a high-throughput experimentally validated regulatory interaction discovery strategy plus manual curation. A typical experimental pipeline proceeds as follows (Figure 3): 1) A regulator is selected and its encoding gene is knocked out. 2) Next, a gene expression analysis is carried out. 3) Then, the promoter regions of genes significantly up- or down-regulated are scanned using MEME/MAST (Bailey et al., 2006) to identify putative binding sites. 4) Finally, electrophoretic mobility shift assays (DNA band shift assays) are used to experimentally validate the putative DNA binding sites. The experimentally validated dataset was expanded by a review of recent literature. We then modeled RNs as directed graphs where nodes represent genes and directed edges stand for regulatory interactions. We want to highlight that the experimental strategy depicted here is not the only high-throughput strategy, but it is the most often used during the last years. Alternative high-throughput strategies are ChIP-chip and ChIP-seq (Jungwirth et al., 2013; Toyoda et al., 2011).

Evidences with varying degrees of confidence support regulatory interactions. Some evidences (e.g., footprinting and electrophoretic mobility shift assays) are strong and directly show that the regulator is binding to the upstream region of the regulated gene, whereas weak evidences just suggest a hypothetical DNA binding site (e.g. bioinformatics predictions) or a possible indirect effect (e.g. gene expression experiments). Weakly supported interactions that eventually turn out to be spurious act like noise affecting the conclusions drawn from RN analyses and mathematical models disregarding this confounding factor (Babtie et al., 2014).

We therefore created two models of the *C. glutamicum* RN: all-evidences (containing both weak and strong supported interactions) and strongly-supported (only accounting for strongly supported evidences). These two RN models were input into the automatized pipeline of Abasy Atlas to obtain a full characterization of the RN models. Most of the predictions here discussed for both of the *C. glutamicum* RN models are available online in Abasy Atlas. There, the RN models are also available for download. An alternative approach to carry out network analyses on the *C. glutamicum* RN is by using corresponding Cytoscape plug-ins such as CoryneRegNetLoader (Baumbach and Apeltsin, 2008) and others for network modularity, hierarchy and centrality analyses (http://apps.cytoscape.org/).

Nowadays, there are no models estimating the total number of regulatory interactions in a particular RN, we therefore evaluated the completeness of each RN model in terms of its percentage of genomic coverage. The all-evidences and strongly-supported RN models are the most complete reconstructions of the *C. glutamicum* RN with a genomic coverage of 73% and 71% (Table 1), respectively. It is an increase of roughly 50% in genomic coverage regarding a previous version of the *C. glutamicum* RN model reconstructed in 2011. This positions our RN models as the third and sixth most complete RN models in Abasy Atlas.

## 4 Global organizational properties, systems-level features and functional architecture

The Abasy Atlas pipeline is responsible for mapping gene symbols into canonical gene names, collecting accession IDs for cross-linking and gene symbols for searching in Abasy Atlas, computing systems and systems-level elements by implementing the NDA, computing network global properties, and annotating each system identified by the NDA.

### 4.1 Global organizational properties

The RN models here reported show an increase of more than 220% in the number of regulatory interactions regarding the 2011 RN model. The number of weakly connected components decreased from 25 (2011 RN model) to three in both novel RN models, showing a more cohesive RN. The average clustering coefficient slightly increased from 0.181 to 0.217 and 0.188 showing that modularity does not increased to a large extend in the novel RN models. The average shortest path length also decreased from 3.53 to 2.20 and 2.14 for all-evidences and strongly-supported, respectively, in agreement with a network having a small-world property.

### 4.1.1 Network diameter and hierarchical layers in a top-down pyramidal hierarchy

The network diameter (the longest of all the calculated shortest paths in a network) decreased from nine (2011 RN model) to seven and six for the all-evidences and strongly-supported RN models, respectively, which is in good agreement with the diameter of the strong-supported RN models of *E. coli* K-12 MG1655 and *B. subtilis* 168. This latter result has important implications regarding the number of hierarchical layers in an acyclic RN organized into a pyramidal top-down hierarchy as the one described by (Ma et al., 2004a). We found that in such pyramidal top-down hierarchy, the number of hierarchical layers equals the diameter of the network. Therefore, there is also a decrease in the number of hierarchical layers in the all-evidences and strongly-supported RN models regarding the 2011 RN model. Additionally, the number of hierarchical layers in *C. glutamicum* is in good agreement with those of *E. coli* and *B. subtilis*, a result updating a previous report claiming that *C. glutamicum* has a lower number of hierarchical layers than *E. coli* (Schröder and Tauch, 2010).

A note further clarifying this point: An acyclic network is indeed a prerequisite to apply the top-down approach described by (Ma et al., 2004a) as in the top-down hierarchy all the regulatory links are downward. The top-down approach places nodes into layers so that a gene is only regulated by the upper layers and regulates the lower layers. On the contrary, if a non-acyclic network is used a paradox arises as we will discuss in section 4.2.2. The way the top-down approach places nodes into hierarchical layers causes that the two most distant nodes (those involved in the network diameter) are located in the extreme hierarchical layers. Thus accounting for the fact that the number of hierarchical layers equals the diameter of the network.

### 4.1.2 Power-law behaviors truly govern the connectivity and clustering coefficient distributions

Despite the suggestion by Barabasi and colleagues (Barabasi and Oltvai, 2004), power laws governing global network distributions have been controversial and even regarded as a myth or dogma (Lima-Mendez and van Helden, 2009). Among the global properties here computed for a RN model, we computed the connectivity, out-connectivity and clustering coefficient distributions. Commonly these distributions are fitted to a power law by ordinary least squares, which is a method that is highly sensitive to outliers in data. Actually, it has been controversial whether using ordinary least squares is an adequate methodology to detect and characterize power-laws in biological networks, and the criticism has even cast doubts in the utility of global properties to study these complex networks (Lima-Mendez and van Helden, 2009). To overcome these issues, we here used two approaches to identify and characterize power laws.

First, we fitted these distributions by robust linear regression of log-log-transformed complementary cumulative data with Huber's T as M-estimator (maximum likelihood-type estimator) to overcome the negative effect of outliers. Second, we used the methodology developed by Clauset et al. (Clauset et al., 2009) as follows: 1) we used maximum likelihood to fit each RN model connectivity data to a power law and alternative heavy-tailed distributions. 2) We then used Kolmogorov-Smirnov to test the goodness of fit of data to each statistical model produced by fitting. 3) Finally, we directly compared power-law fits against alternative models by using likelihood ratio tests. The latter approach is only applicable to probability distributions such as the connectivity distribution, but the clustering coefficient distribution does not meet this requirement. Therefore, we characterized clustering coefficient distributions only by the former approach (robust linear regression).

Robust linear regression shows that connectivity and clustering coefficient distributions follow power laws (Figure 4), which improve showing a better goodness of fit ($R^2$) in both of the RN models here reported regarding the 2011 RN model (from 0.98 to 0.99 for both $P(k)$; from 0.69 to 0.94 and 0.96 for all-evidences and strongly-supported $C(k)$, respectively). To evaluate the robust-linear-regression characterizations, we contrasted the connectivity and out-connectivity distributions obtained against their versions fitted by maximum likelihood across the three RN models (Table 2). We found that power-law exponents obtained by robust linear regression are in good agreement with those estimated by maximum likelihood. The goodness of fit (Kolmogorov-Smirnov $D$ statistic) of each RN model is better for a power law than alternative distributions. Loglikelihood ratio tests ruled out exponential distribution in all the RN models. While stretched exponential and log-normal distributions cannot be ruled out by loglikelihood ratio test, they are excluded because of their goodness of fit, except for the 2011 RN model. Actually, this RN model produce the worse power laws, for instance, in the out-conectivity data of the 2011 RN model is evident a tie between power-law, stretched exponential and log-normal distributions. Indeed, it also provides evidence contradicting the claim that power laws could emerge due to network incompleteness (Lima-Mendez and van Helden, 2009), as a study only focused in the most incomplete 2011 RN model does not provide enough support to conclude that power laws govern connectivity distributions. In conclusion, the novel RN models here reported show that the power law is not a myth in network biology as previously claimed, and that robust linear regression is a very good approximation to characterize them in RN.

### 4.2 Circuit motifs, feedback loops and autoregulation

#### 4.2.1 Circuit motifs

A circuit or topological motif is defined as a statistically over-represented pattern of interconnected nodes and links (subgraphs) in a complex network (Milo et al., 2002). Two principal RN motifs have been found in RNs: the feed-forward motif (FF) and the bi-fan motif (Shen-Orr et al., 2002). FFs are three-node motifs comprising two regulatory genes and one target gene (A → B, B → C, A → C), and bi-fans involve two regulatory genes cross-regulating two target genes (A → C, A → D, B → C, B → D). FFs perform various dynamical roles including sign-sensitive delay, persistence detection, response acceleration and pulse generation (Alon, 2007; Macia et al., 2009; Wall et al., 2005). Some studies have even suggested that the dynamical roles played by FFs could explain why they have been selected in RNs (Alon, 2007; Mangan and Alon, 2003). On the contrary, other studies have shown that the overabundance of bi-fans does not correlate with any specific functional role (Ingram et al., 2006). Recent evidence suggests that motifs in RNs could be a by-product resulting from network organization and evolution (Cordero and Hogeweg, 2006; Freyre-Gonzalez et al., 2008; Mazurie et al., 2005; Sole and Valverde, 2006). Actually, some studies have highlighted that FFs play an important organizational (Freyre-Gonzalez et al., 2008) role on the functional architecture. Hence, we here only focus our attention on three-nodes motifs as the FF.

We find that the common motif in the all-evidences and-strongly-supported RN models is not the FF but an alternative version consisting of a two-node feedback loop circuit between the regulatory nodes, hereinafter called the complex feed-forward motif (CFF) (163 and 104 instances, respectively). We also found this CFF in the *E. coli* K-12 MG1655 and *B. subtilis* 168 RN models. The number of FFs and CFFs found in each bacterial RN model is available in Abasy Atlas (Ibarra-Arellano et al., 2016). Indeed, the CFF is widely distributed in bacteria as roughly the 22% of the RN models in Abasy Atlas has > 100 CFFs. We only found the classical FF (587 instances) in the strongly-supported RN model. These results contrast with those found in the 2011 RN model where the only motif found was the FF (404 instances).

### 4.2.2 Feedback loops and autoregulation

Feedback loops (FBLs) are circular regulatory interactions necessary to give rise to different cellular behaviors, such as homeostasis and differentiation (Kaern et al., 2005; Smits et al., 2006; Thomas, 1998; Thomas and Kaufman, 2001). Hereinafter, we distinguished between autoregulation and FBL. By FBL, we mean those circular interactions involving two or more genes, disregarding autoregulation. In graph theory, an *n*-cycle is a FBL comprising *n* nodes or genes. An autoregulation or self-loop is a special case of a FBL ($n = 1$) and is discussed towards the end of this section.

Disregarding autoregulation, some studies have claimed that RNs are acyclic (Balazsi et al., 2005; Ma et al., 2004a), whereas other has suggested that FBLs were not previously identified because the genes composing a FBL are located within the same operon (Ma et al., 2004b). Indeed, the idea that a RN functional architecture is acyclic and pyramidal is widely disseminated in the literature (Balazsi et al., 2005; Bhardwaj et al., 2010a; Bhardwaj et al., 2010b; Janga et al., 2009; Ma et al., 2004a; Ma et al., 2004b; Martinez-Antonio and Collado-Vides, 2003; Rodriguez-Caso et al., 2009; Yan et al., 2010; Yu and Gerstein, 2006). In contrast, other studies have been shown that even disregarding autoregulation the *E. coli* and *B. subtilis* RNs are non-acyclic, and genes not encoded within the same operon generally compose FBLs bridging different hierarchical layers of the functional architecture (Freyre-Gonzalez et al., 2008; Freyre-Gonzalez et al., 2012).

The sole presence of FBLs, whether they are statistically over-represented or not, poses a paradox to infer the hierarchy governing a RN: given the circular nature of their interactions, what nodes should be placed in a higher hierarchical layer? Neglecting the presence of FBLs can lead to biologically incorrect hierarchies placing pleiotropic regulators in lower hierarchical layers. The NDA solves this paradox by disaggregating genes into classes of systems-level elements with different hierarchical levels (Figure 2).

A full enumeration of the FBLs present in the *C. glutamicum* RN models found that the number of FBLs slightly growth from two (2-cycles = 2; *z*-score = 1.24) in the 2011 RN model to eight (2-cycles = 6 and 3-cycles = 2; *z*-score = 0.48) and five (2-cycles = 4 and 3-cycles = 1; *z*-score = 1.60) in the all-evidences and strongly-supported RN models, respectively. We found that FBL are not statistically over-represented (*z*-score < 2) in the RN models when we contrasted the number of FBLs in each RN model against the expected in 1,000 randomly rewired networks conserving the in-degree and out-degree distributions. Despite FBL are not over-represented, we observed that the 38% (3/8) and 40% (2/5) of the FBLs in the all-evidences and strongly-supported RN models involve at least one global regulator and bridge different hierarchical layers in the functional architecture. Additionally, we found that none of the FBLs is composed of genes encoded within the same operon. All these evidences support a model where RNs are non-acyclic and FBLs, despite not being statistically over-represented, play an important role by cross-regulating the coordination and processing layers.

As stated above, a trivial case of FBL is the autoregulation or self-loop where a single gene controls its own expression. We found that the fraction of autoregulations slightly increased from 42% in the 2011 RN model to 51% and 50% in the all-evidences and strongly-supported RN models, respectively. Autoregulations (in 2011, all-evidences and strongly-supported RN models) are mostly negative (85%, 85% and 82%) with some being positive (15%, 12% and 13%) and a small percentage dual or unknown (0%, 3% and 5%). This is in agreement with a trend previously reported for *C. glutamicum* (Schröder and Tauch, 2010), *E. coli* (Perez-Rueda and Collado-Vides, 2000; Thieffry et al., 1998) and *B. subtilis* (Moreno-Campuzano et al., 2006).

### 4.3 Functional architecture and systems-level elements

The distribution of the four classes of systems-level elements predicted for the all-evidences and strongly-supported *C. glutamicum* RN models reported here is summarized in Table 1.

The Abasy Atlas pipeline annotated each NDA-predicted system and the clusters of intermodular and basal machinery genes by computing their functional enrichment. Functional enrichment analysis was carried out by computing *p*-values for each gene ontology (GO) term in the cluster of genes using a hypergeometric distribution. We considered only GO terms in the biological process namespace because they reflect physiological processes. We then corrected *p*-values for multitests producing *q*-values by controlling the false discovery rate at level 0.05. We selected UniProt GO annotation as the functional classification schema for systems annotation because it is actively and manually curated (Huntley et al., 2015) and provides a shared vocabulary among organisms and databases (Ashburner et al., 2000).

#### 4.3.1 Coordination layer: Global regulators

The global regulators compose the coordinating layer in the functional architecture predicted by the NDA. They are identified in the NDA by defining an equilibrium point (κ-value) between two apparently contradictory behaviors occurring in hierarchical-modular networks: hubness and modularity. In these networks, modularity is inversely proportional to hubness (Figure 1). The κ-value is defined as the connectivity value where the variation of the clustering coefficient (modularity) equals the variation of out-connectivity (hubness) but with the opposite sign ($dC/dk_{out} = -1$) (Freyre-Gonzalez et al., 2008; Freyre-Gonzalez et al., 2012). This is a network-depending parameter that allows the identification of global regulators as those genes with connectivity > κ.

In a previous *C. glutamicum* RN model reconstructed in 2011 and reported in the first version of Abasy Atlas, we found 19 global regulators (2.6%, κ-value = 11): *cg2092* (*sigA*), *glxR*, *cg0876* (*sigH*), *dtxR*, *ramA*, *ramB*, *lexA*, *cg3253* (*mcbR*), *cg2115* (*sugR*), *cg0986* (*amtR*), *cg0862* (*mtrA*), *cg3247* (*hrrA*), *ripA*, *cg2888* (*phoR*), *cg3420* (*sigM*), *cg0156* (*cysR*), *cg1324* (*rosR*), *cg2102* (*sigB*) and *cg0196* (*iolR*) (Ibarra-Arellano et al., 2016). These global regulators comprise four out of the seven sigma factors encoded in the genome of *C. glutamicum* (Patek and Nesvera, 2011). All of these regulators have been jointly described as global regulators (Schröder and Tauch, 2010) except for *cg0862* (*mtrA*), *cg3247* (*hrrA*), *ripA*, *cg3420* (*sigM*), *cg0156* (*cysR*), *cg1324* (*rosR*) and *cg0196* (*iolR*). Nevertheless, *cg3247* (*hrrA*) has also been individually reported as global regulator (Frunzke et al., 2011). In our set of global regulators we only missed *arnR* (*cg1340*) that has been reported as a global regulator (Schröder and Tauch, 2010), but it is a modular gene belonging to a system annotated as "nitrate metabolic process" according to the NDA.

Interestingly, in both of the new *C. glutamicum* RN models here analyzed, we identified only four global regulators (0.2%, κ-value$_{all\text{-}evidences}$ = 58 and κ-value$_{strongly\text{-}supported}$ = 57): *cg2092* (*sigA*), *glxR*, *cg0876* (*sigH*) and *dtxR*. All these global regulators are included in the predictions of the 2011 RN model. One of the main sources of regulatory interactions in the herein reported RN models is the addition of 1,848 *sigA*-dependent strongly-supported interactions discovered by RNA-seq (Pfeifer-Sancar et al., 2013). These account for the 55% of the total interactions in the all-evidences RN model. We evaluated whether this could be biasing the prediction of global regulators by adding 1,800 new *sigA*-dependent interactions to the 2011 RN model. We found that the number of predicted global regulators decreased from 19 (2.6%) to 10 (0.4%), roughly the same fraction observed in both of the new RN models. In addition, the number of basal machinery genes increased from 405 (54.5%) to 2,089 (82.1%), which is in agreement with the increase observed in the all and strong evidences RN models (1,631 (70.9%) and 1,713 (77.1%), respectively). These evidences support the idea that dramatically increasing the out-connectivity of the main network hub (*sigA*) bias the κ-value as it depends on an equilibrium between out-connectivity and modularity. All these suggest that global regulators prediction will improve as new local interactions will be discovered.

### 4.3.2 Processing layer: Functional systems (modules) and basal machinery

Two classes of systems-level elements comprise the processing layer of a RN: functional systems shaped by modular genes and basal machinery genes. The number of systems in the novel RN models went from 61 to 68 and 56 in the all-evidences and strongly-supported RN models, respectively. The fraction of functionally annotated systems slightly increased from 33% to roughly 40% in both RN models. The low rate of functional annotation is due to the low GO-annotation coverage as only 54% (1,234/2,301) and 53% (1,182/2,223) of the genes in the all-evidences and strongly-supported RN models, respectively, are GO-annotated.

We observed a decrement in the fraction of modular genes from 42.7% (317) to 24.4% (562) and 20.6% (459) and an increment in the fraction of basal machinery genes from 54.5% (405) to 70.9% (1,631) and 77.1% (1,713). This is in agreement with the slight increase in the average clustering coefficient mentioned in section 4.1. We hypothesize that the curation of high-throughput experiments explains the increase in the basal machinery genes and low modularity as many global interactions are added but local interactions shaping systems remain to be discovered, thus highlighting the importance to carry out more experiments aimed to unveil local interactions.

One of the prominent features in Abasy Atlas is that it goes beyond genomic functional associations providing systems-level functional associations. It enables the identification of the function of genes whose annotation by homology analysis is not possible. We found 93 and 73 modular genes in the all-evidences and strongly-supported RN models, respectively, annotated as hypothetical or with missing annotation. The 66% (61/93) and 53% (39/73), respectively, of these modular genes actually belong to functionally annotated systems providing a way to assign functional annotations to these genes by using a systems-level guilt-by-association strategy: if gene G belongs to a system with function X, then G is also involved in X.

Basal machinery genes are those exclusively regulated by global regulators. It has been previously reported that these genes mostly comprise elements of the basal cell machinery (e.g., tRNAs and their charging enzymes, DNA and RNA polymerases, ribosomal elements, DNA repair/packing/segregation/methylation) and poorly studied systems (Freyre-Gonzalez et al., 2012). Here, we found two common GO terms enriched in both of the *C. glutamicum* RN models: "translation" (GO:0006412) and "cell wall organization" (GO:0071555). These are the only functional terms enriched in the strongly-supported RN model. However, the all-evidences RN model adds others such as "cell cycle" (GO:0007049), "regulation of cell shape" (GO:0008360), "peptidoglycan biosynthetic process" (GO:0009252), and "cell division" (GO:0051301). All of these functions are closely related to the basal cell machinery thus supporting that these class of genes are not due to an artifact of the NDA or network incompleteness.

### 4.3.3 Integration layer: Intermodular genes

Intermodular genes are a novel system-level element first identified by the NDA (Freyre-Gonzalez et al., 2008; Freyre-Gonzalez and Trevino-Quintanilla, 2010; Freyre-Gonzalez et al., 2012). They compose the integrative layer of a RN by combining, at the promoter level, disparate physiological responses from different modules to achieve a unified response (Figure 2).

We found an increase in the number of intermodular genes in both of the new RN models here reported against the 2011 RN model. The number of intermodular genes is 104 (4.5%) and 47 (2.1%) in the all-evidences and strongly-supported RN models, respectively. In contrast with only two (0.3%) intermodular genes identified in the 2011 RN model. The fraction of intermodular genes found in the new RN models are consistent with those found in the RN models of *E. coli* K-12 MG1655 and *B. subtilis* 168 (Freyre-Gonzalez et al., 2012; Ibarra-Arellano et al., 2016) (Table 1).

A functional enrichment analysis shows that "pentose-phosphate shunt" (GO:0006098) and "nitrate metabolic process" (GO:0042126) are the common GO terms for the intermodular genes identified in both RN models. The all-evidences RN model is also enriched with "glutamate biosynthetic process" (GO:0006537), "pyridoxal phosphate biosynthetic process" (GO:0042823) and "urea catabolic process" (GO:0043419). On the other hand, the strongly-supported RN model is also enriched with "acetyl-CoA biosynthetic process" (GO:0006085) and "leucine biosynthetic process" (GO:0009098).

An interesting example of an intermodular gene in the strongly-supported RN model is *hemH* (Figure 5). This gene encodes a ferrochelatase responsible for protoporphyrin metalation, whose expression is repressed by Cg2109 (OxyR) in module 1.2 (cellular iron ion homeostasis) and Cg3247 (HrrA) in module 1.3 (protoporphyrinogen IX biosynthetic process). The cross-regulation of *hemH* enables the cell to maintain a balance between the production of heme B (protoheme IX) and keeping the iron homeostasis under low-iron conditions. This example highlights the integrative role of intermodular genes. A couple of also interesting and more complexly regulated examples of intermodular genes are *acn* (Figure 6) and *cg0446* (*sdhA*) (Figure 7) in the strongly-supported RN model. These genes were foreseen as intermodular genes by a couple of previous studies based solely on their complex regulatory pattern (Brinkrolf et al., 2010; Schröder and Tauch, 2010).

### 4.3.4 Effects of network incompleteness on the robustness and predictions accuracy of the natural decomposition approach

Current RN models are largely incomplete (Rottger et al., 2012), and this could potentially bias our results. RN models have two potential sources of incompleteness: genes (nodes) and interactions (edges). As mentioned above, the *C. glutamicum* RN models reconstructed here have a high genomic coverage (completeness in the set of genes) of 73% and 71% for the all-evidences and strongly-supported RN models, respectively (Table 1). On the other hand, according to a statistical estimation, our RN models roughly comprise the 48% and 35% of the total regulatory interactions (completeness in the set of interactions) for the all-evidences and strongly-supported RN models, respectively. Thus, the main source of incompleteness in our RN models comes from the set of interactions.

To assess the robustness of the NDA predictions we carry out a sensitivity analysis by computing the overlap among predictions across a diverse range of randomly perturbed RN models derived from the strongly-supported *C. glutamicum* RN model. We selected the strongly-supported RN model to minimize the external noise in our sensitivity analysis. We separately assessed the effect of the incompleteness in the set of genes or interactions by randomly removing a fraction of genes or interactions, respectively (Figure 8).

We observed that the global regulators class is the most robust to network incompleteness, whereas the intermodular genes class is the most sensitive. The latter being almost so sensitive as the modular genes and basal machinery under random genes removal. It is fair to say that the sensitivity of the intermodular genes class is virtually invariant to the type of perturbation (genes or interactions removal), being slightly greater under genes removal.

We then discarded the global regulators and intermodular genes classes as outliers and focused our attention on the modular genes and basal machinery classes. We found that the NDA is highly robust to incompleteness in the set of interactions, as we need to remove at least 70% of interactions in both types of perturbations to observe that the overlap falls below 0.5. On the other hand, the NDA is more sensitive to incompleteness in the set of genes, as the overlap starts to fall below 0.5 after removing only 30% of genes in both perturbed RN models. These results in conjunction with the high genomic coverage of the RN models herein reconstructed show that the NDA predictions reported in this study are highly robust to RN models incompleteness.

## 5   Conclusions

The *C. glutamicum* RN models here constructed, and available for download in Abasy Atlas, provide the most complete reconstruction of the *C. glutamicum* RN to date. Our RN models position as the third and sixth most complete RN models available in Abasy Atlas after the meta-curations of *Mycobacterium tuberculosis* H37Rv and *B. subtilis* 168. Using the framework provided by Abasy Atlas, we comprehensively explored the global organizational properties, systems-level features and functional architecture of the *C. glutamicum* RN. Our analyses show the tremendous power of Abasy Atlas to study the functional organization of RNs. Using state-of-the-art methodologies, we shed new light on long-standing issues. We prove that power-law behaviors truly govern the connectivity and clustering coefficient distributions. This is key as a part of the NDA, the κ-value computation, is based on the clustering coefficient distribution. The high predictive power of the NDA for unveiling the functional architectures of bacterial RNs proves that global distributions are useful to study RN organization. We discuss that a novel circuit motif, the CFF, was found in the *C. glutamicum* RN. We highlighted the importance of FBLs for the functional architecture beyond whether they are statistically over-represented or not in the network. The presence of FBLs shows that it is inadequate to apply the previously reported top-down approach to infer the pyramidal hierarchy of a RN because they bridge different layers of the functional architecture. Inferring a pyramidal hierarchy is also inadequate as it disregards the presence of intermodular genes shaping the integration layer. These pieces of evidence all together further support a diamond-shaped, three-layered hierarchy exhibiting some feedback between processing and coordination layers, which is shaped by four classes of systems-level elements: global regulators, locally autonomous modules, basal machinery and intermodular genes. Previous work has shown that evolutionary rewiring converge into the same functional architecture governing highly plastic RNs (Freyre-Gonzalez et al., 2012). This convergence shapes an organizational landscape. This landscape represents the space of topological properties delimiting possible bacterial RNs. We now require further studies to explore what are the rules governing the evolutionary rewiring of RNs enabling the convergence to a well-delimited organizational landscape. Nevertheless, the number and size of the available curated regulatory data sets are factors limiting the diversity of RN models to study the general principles governing the evolutionary rewiring of RNs. To overcome this limitation, it is important to adopt pipelines to transfer regulatory knowledge among organisms as has been previously described (Baumbach, 2010; Novichkov et al., 2013).

## 6   Availability of supporting data

The data sets supporting the results of this article are available for download in Abasy Atlas (http://abasy.ccg.unam.mx).

## 7   List of abbreviations

RN, regulatory network; NDA, natural decomposition approach; FF, feed-forward motif; CFF, complex feed-forward motif; FBL, feedback loop; GO, gene ontology

## 8   Acknowledgements

We thank Adrian I. Campos-González for technical support on the sensitivity analysis of the NDA predictions. This work was supported by the Programa de Apoyo a Proyectos de Investigación e Innovación Tecnológica (PAPIIT-UNAM) [IA200614 and IA200616 to JAF-G].

Yan, K.K., Fang, G., Bhardwaj, N., Alexander, R.P., Gerstein, M., (2010) Comparing genomes to computer operating systems in terms of the topology and evolution of their regulatory control networks. Proc Natl Acad Sci U S A 107, 9186-9191.

Yu, H., Gerstein, M., (2006) Genomic analysis of the hierarchical structure of regulatory networks. Proc Natl Acad Sci U S A 103, 14724-14731.

## 10 Figures

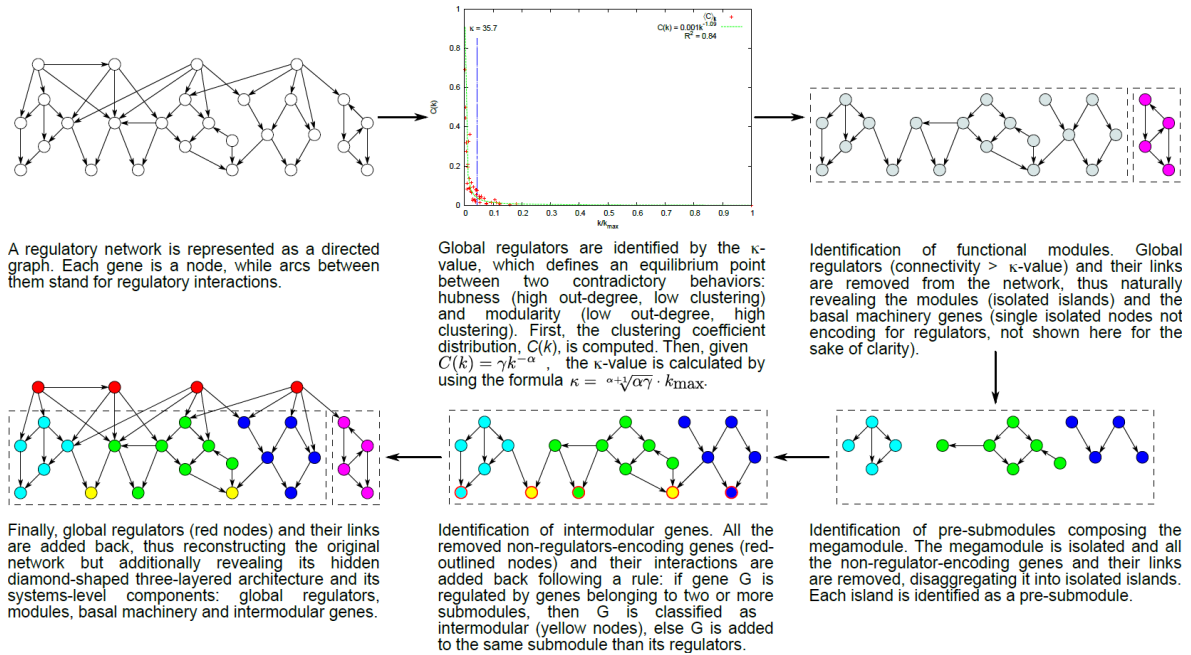

Figure 1. The natural decomposition approach. Figure taken from (Ibarra-Arellano et al., 2016).

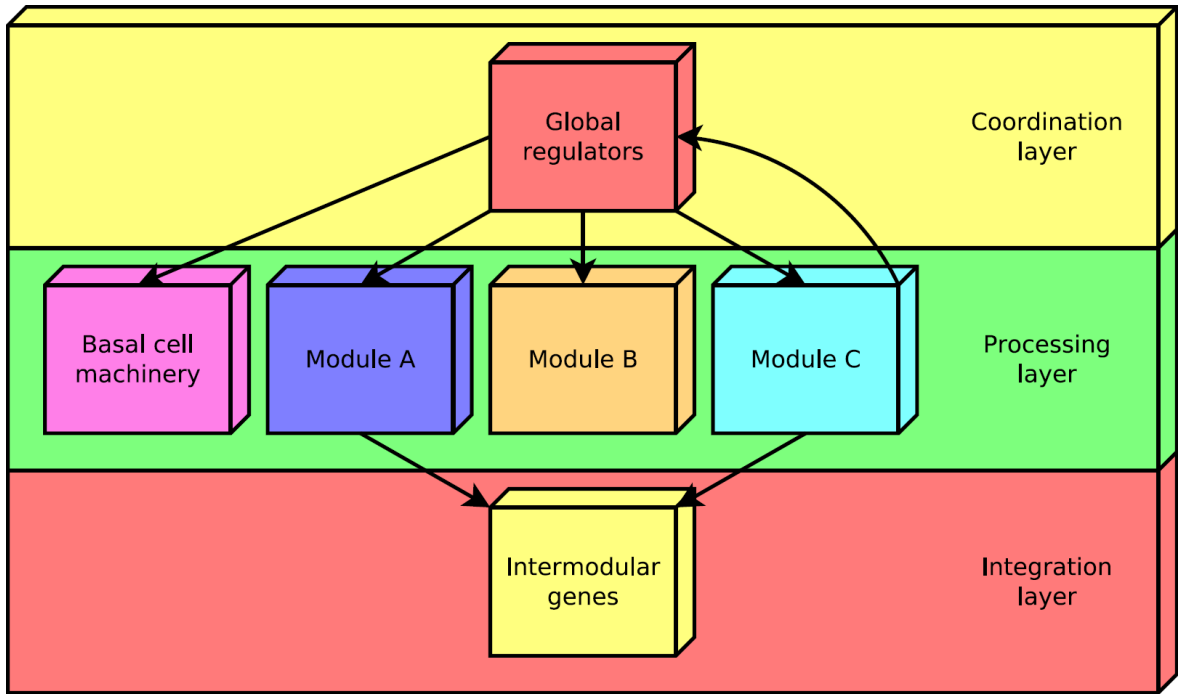

Figure 2. Common functional architecture identified by the natural decomposition approach across bacteria. The functional architecture unveiled by the natural decomposition approach is a diamond-shaped, three-layered hierarchy exhibiting some feedback between processing and coordination layers, which is shaped by four classes of systems-level elements: global regulators, locally autonomous modules, basal machinery and intermodular genes. Figure modified from (Ibarra-Arellano et al., 2016).

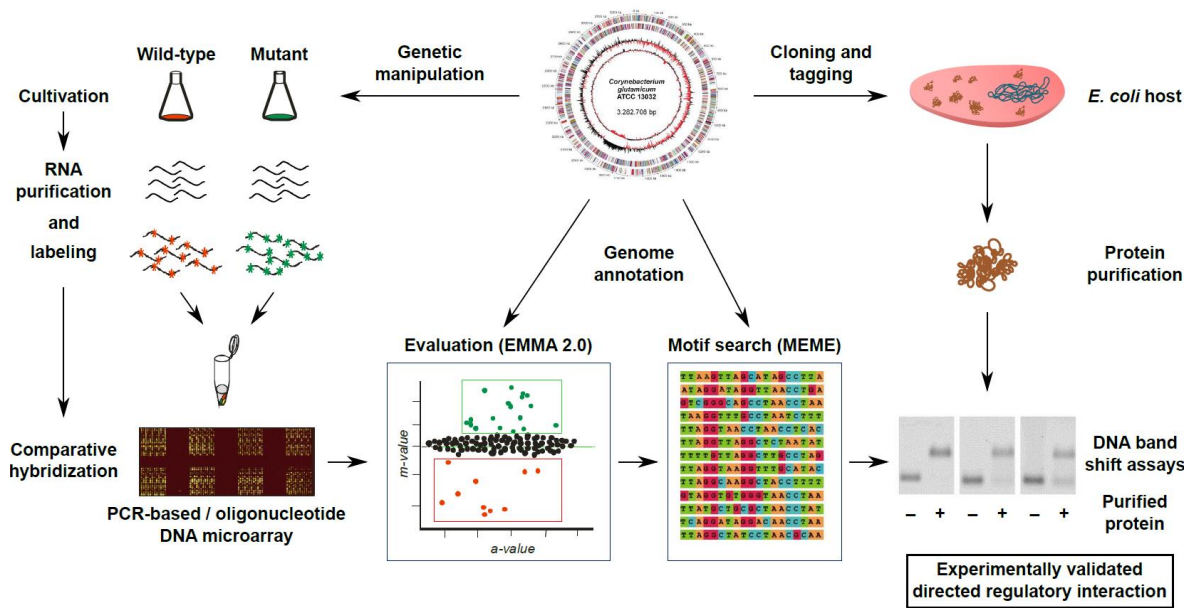

Figure 3. Methodology used for the discovery of experimentally validated regulatory interactions in *C. glutamicum*.

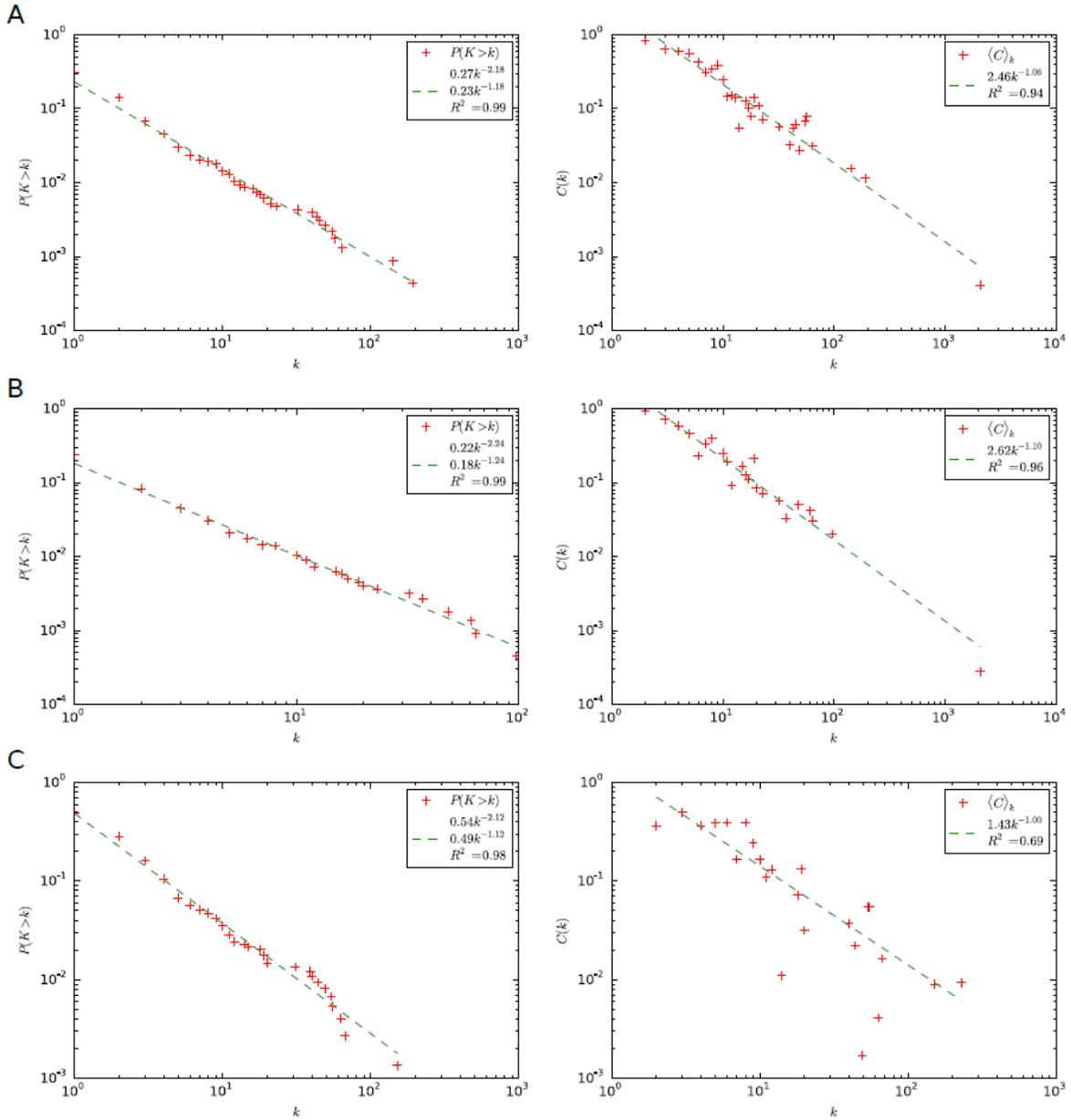

Figure 4. Connectivity ($P(k)$) and clustering coefficient ($C(k)$) distributions of the (A) all-evidences, (B) strongly-supported and (C) 2011 *C. glutamicum* RN models obtained by robust linear regression.

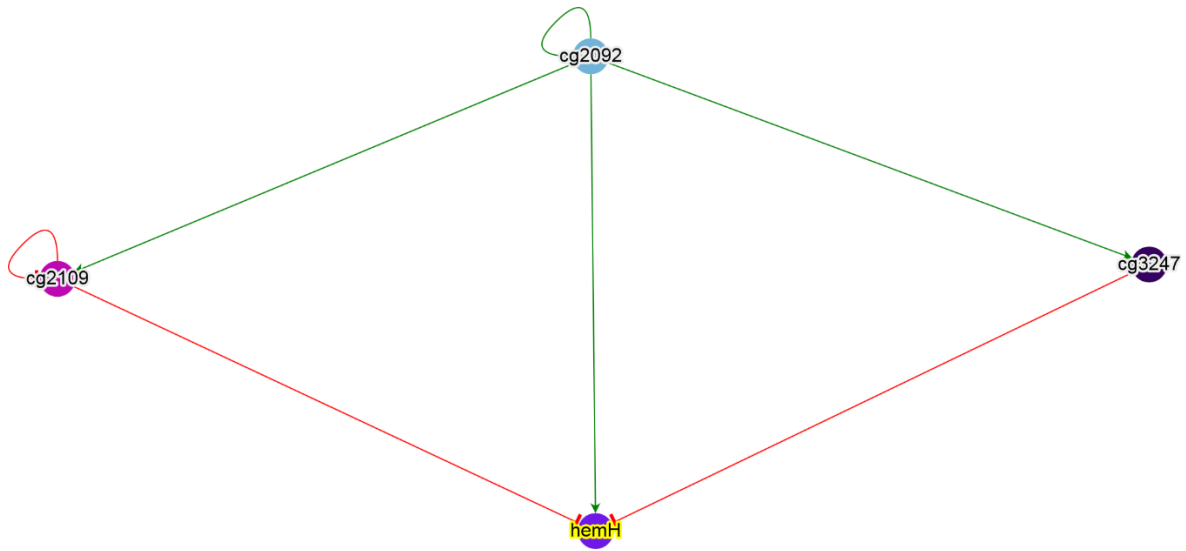

Figure 5. The intermodular gene *hemH*, highlighted in yellow, encodes a ferrochelatase responsible for protoporphyrin metalation, which is regulated by the product of genes *cg2109* (*oxyR*) in module 1.2 (cellular iron ion homeostasis) and *cg3247* (*hrrA*) in module 1.3 (protoporphyrinogen IX biosynthetic process). The gene *cg2092* (*sigA*) encodes a common global regulator. Red lines with T-shaped arrowhead represent repressions whereas green lines with a normal arrowhead stand for activations. The figure was generated with Abasy Atlas (Ibarra-Arellano et al., 2016).

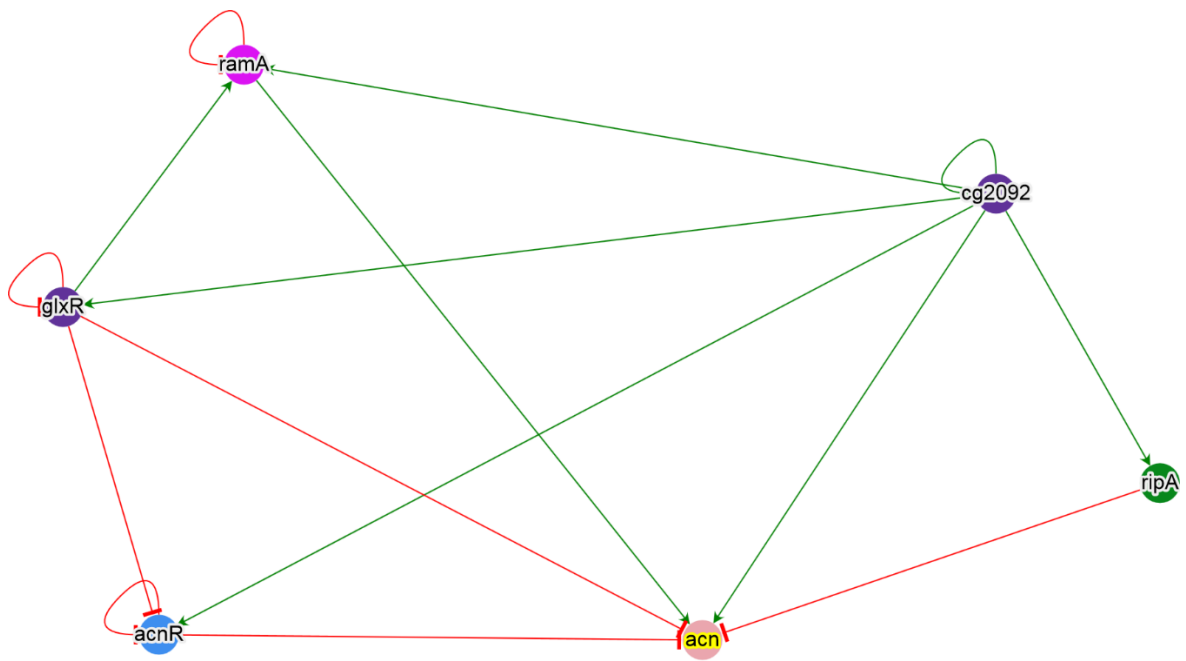

Figure 6. The intermodular gene *acn*, highlighted in yellow, encodes an aconitate hydratase and is complexly regulated by two global regulators (purple nodes: Cg2092 (SigA) and GlxR) and three modular regulators: AcnR in module 1.6 (unknown function), RamA in module 1.1 (cellular carbohydrate metabolism) and RipA in module 1.16 (catechol-containing compound metabolism).

Red lines with T-shaped arrowheads represent repressions whereas green lines with normal arrowheads stand for activations. The figure was generated with Abasy Atlas (Ibarra-Arellano et al., 2016).

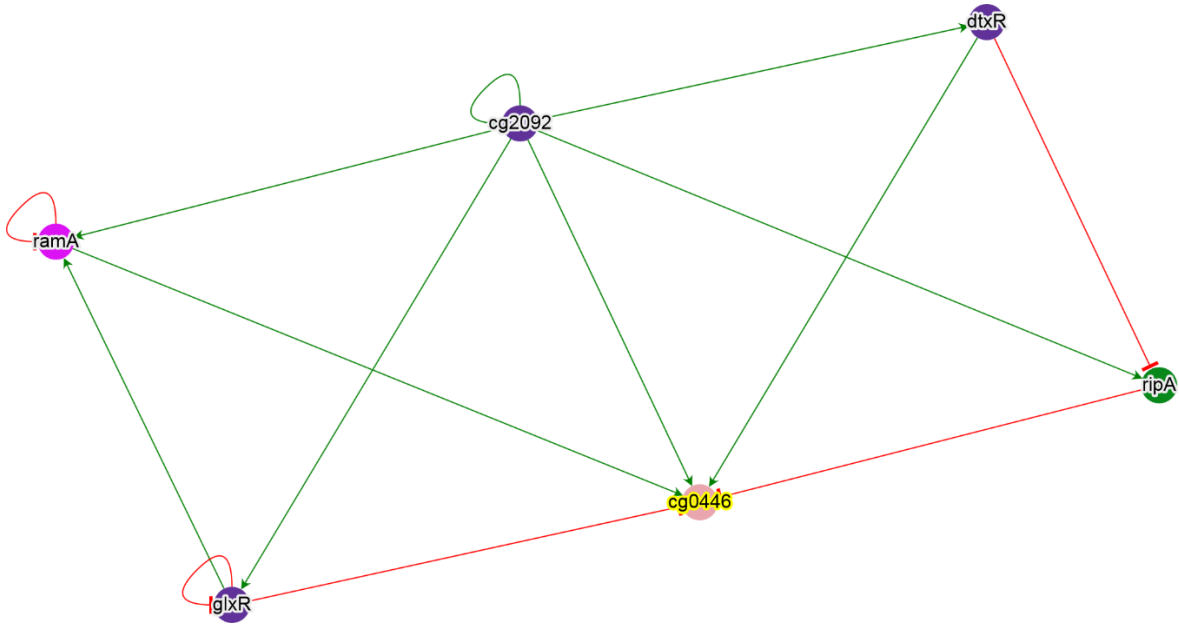

Figure 7. The intermodular gene *cg0446* (*sdhA*), highlighted in yellow, encodes a succinate dehydrogenase flavoprotein subunit and is complexly regulated by three global regulators (purple nodes: Cg2092 (SigA), GlxR and DtxR) and two modular regulators: RamA in module 1.1 (cellular carbohydrate metabolism) and RipA in module 1.16 (catechol-containing compound metabolism). Red lines with T-shaped arrowheads represent repressions whereas green lines with normal arrowheads stand for activations. The figure was generated with Abasy Atlas (Ibarra-Arellano et al., 2016).

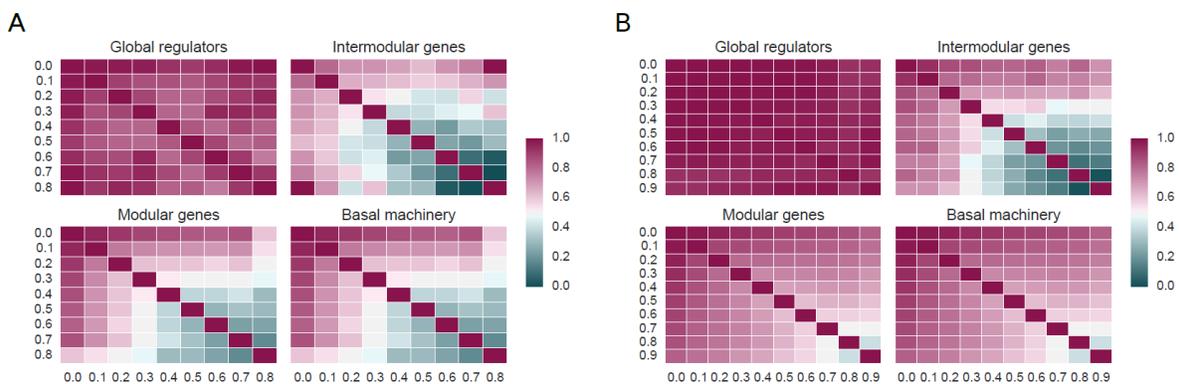

Figure 8. Sensitivity analysis of the natural decomposition approach predictions by randomly removing (A) genes and (B) interactions from the strongly-supported *C. glutamicum* RN model. Axes show the fraction of (A) genes and (B) interactions removed. Each cell of both heat maps is from averaging the overlap coefficient among predictions over 100 realizations of the particular random perturbation.

## 11 Tables

Table 1 – Comparison ordered by genomic coverage among the systems-level compositions of the *C. glutamicum* RN models reconstructed here (first two rows) in contrast to *E. coli* and *B. subtilis* strongly-supported RN models available in Abasy Atlas (Ibarra-Arellano et al., 2016) and the 2011 *C. glutamicum* RN model.

| Regulatory network model | Genomic coverage | Global regulators[a] | Systems (modular genes)[a] | Basal machinery genes[a] | Intermodular genes[a] |
|---|---|---|---|---|---|
| *C. glutamicum* ATCC 13032 (all evidences)[b] | 73.3% (2301) | 4 (0.17%) | 68 (24.4%) | 1631 (70.9%) | 104 (4.52%) |
| *C. glutamicum* ATCC 13032 (strong evidences)[b] | 70.8% (2223) | 4 (0.18%) | 56 (20.6%) | 1713 (77.1%) | 47 (2.11%) |
| *E. coli* K-12 MG1655 (strong evidences) | 42.0% (1889) | 14 (0.74%) | 101 (46.3%) | 895 (47.4%) | 106 (5.61%) |
| *B. subtilis* 168 (strong evidences) | 31.8% (1408) | 15 (1.07%) | 81 (51.5%) | 615 (43.7%) | 53 (3.76%) |
| *C. glutamicum* ATCC 13032 (2011)[b] | 23.6% (741) | 19 (2.56%) | 61 (42.7%) | 405 (54.5%) | 2 (0.27%) |

[a]Percentages are relative to the total number of genes in each RN.

[b]RN models constructed in this work.

Table 2 – Evaluation of power-law behavior in the connectivity ($k$) and out-connectivity ($k_{out}$) data of *C. glutamium* RN models against alternative distributions by using robust linear regression (RLR), maximum likelihood (ML), Kolmogorov-Smirnov and likelihood ratio tests. Minimum values per row of goodness of fit ($D$ statistic) are denoted in **bold**. Loglikelihood ratio statistics (LLR) supported by a statistical significant *p*-value (< 0.05) are highlighted in **bold** too. The last column lists our conclusion of the statistical support for the power-law hypothesis for each RN model. "Poor" indicates that power law is a good fit but alternative distributions are also possible explanations. "Good" indicates that power law is the best fit and alternatives are not plausible.

| *C. glutamicum* regulatory network model | Power-law exponent RLR | Power-law exponent ML | Power-law $D$ | Exponential $D$ | Stretched exponential $D$ | Log-normal $D$ | Exponential LLR | Stretched exponential LLR | Log-normal LLR | Support for power law |
|---|---|---|---|---|---|---|---|---|---|---|
| All evidences ($k$) | 2.18 ± 0.03 | 2.12 ± 0.14 | **0.06** | 0.53 | 0.11 | 1.00 | **2.41** | 1.07 | -1.01 | Good |
| Strong evidences ($k$) | 2.24 ± 0.03 | 2.08 ± 0.23 | **0.08** | 0.50 | 0.12 | 1.00 | **3.06** | 1.09 | -0.99 | Good |
| 2011 ($k$) | 2.12 ± 0.03 | 2.04 ± 0.15 | **0.07** | 0.31 | 0.10 | 0.09 | **2.14** | -0.38 | -0.50 | Good |
| All evidences ($k_{out}$) | 1.92 ± 0.03 | 2.06 ± 0.16 | **0.08** | 0.53 | 0.11 | 1.00 | **2.56** | 1.03 | -1.00 | Good |
| Strong evidences ($k_{out}$) | 1.95 ± 0.04 | 1.93 ± 0.23 | **0.08** | 0.52 | 0.10 | 1.00 | **3.15** | 0.80 | -1.00 | Good |
| 2011 ($k_{out}$) | 1.87 ± 0.04 | 1.99 ± 0.16 | **0.10** | 0.29 | **0.10** | **0.10** | 1.74 | -0.58 | -0.56 | Poor |